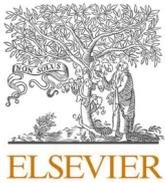
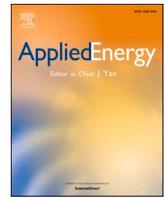

# Health-aware energy management for multiple stack hydrogen fuel cell and battery hybrid systems

Junzhe Shi [a], Ulf Jakob Flø Aarsnes [b], Shengyu Tao [a,*], Ruiting Wang [a], Dagfinn Nærheim [c], Scott Moura [a,*]

[a] *University of California, Berkeley, Department of Civil and Environmental Engineering, 760 Davis Hall, Berkeley, CA 94720, USA*
[b] *NORCE, Tullins gate 2, 0166 Oslo, Norway*
[c] *Corvus Energy, Sandbrekketoppen 30, 5224 Nesttun, Bergen, Norway*

## HIGHLIGHTS

- Developed MIQP-based EMS for FC/battery systems to cut fuel consumption cost and reduce battery and FC degradation.
- Introduced Individual FC Stack Control method to improve system performance.
- Realized significant improvements in computational efficiency over the traditional DP method.

ABSTRACT

Fuel cell (FC)/battery hybrid systems have attracted substantial attention for achieving zero-emissions buses, trucks, ships, and planes. An online energy management system (EMS) is essential for these hybrid systems, it controls energy flow and ensures optimal system performance. Key aspects include fuel efficiency and mitigating FC and battery degradation. This paper proposes a health-aware EMS for FC and battery hybrid systems with multiple FC stacks. The proposed EMS employs mixed integer quadratic programming (MIQP) to control each FC stack in the hybrid system independently, i.e., MIQP-based individual stack control (ISC), with significant fuel cost reductions, FC and battery degradations. The proposed method is compared with classical dynamic programming (DP), with a 2243 times faster computational speed than the DP method while maintaining near-optimal performance. The case study results show that ISC achieves a 64.68 % total cost reduction compared to CSC in the examined scenario, with substantial reductions across key metrics including battery degradation (4 %), hydrogen fuel consumption (22 %), fuel cell idling loss (99 %), and fuel cell load-change loss (41 %)

## 1. Introduction

Hydrogen fuel cells (FC) [1], due to their high energy density, are considered a key alternative for sustainable energy systems [2]. In transportation applications, FC and battery hybrid systems have gained significant attention for the integration of the long-range capability of FCs and the regenerative braking capabilities of batteries [3]. This combination forms an environment-friendly power source that provides high efficiency, extended range, reduced emissions, and enhanced reliability [4] [5].

Most FC and battery hybrid systems consist of a single FC stack [6], and many studies have examined various aspects of single-stack systems, including structural design [7], degradation modeling [8], and energy management strategies [9]. However, many practical applications require higher power output than a single FC stack can provide. For example, a Honda Clarity FC stack delivers a maximum of 103 kW [10]. Meanwhile, city buses require up to 600 kW [11], necessitating for multiple FC stacks to meet higher power demands. Heavy-duty vehicles, such as trucks, and ocean-going vessels, operate under prolonged conditions with rapid transient loads, necessitating multiple FC stacks to satisfy the system response time. Beyond the power, multiple FC stacks provide higher control flexibility, improving overall system operational efficiency and reliability. This can be achieved through Individual Stack

**Nomenclature**

*Vehicle model and system configurations*
| | |
|---|---|
| $m$ | Vehicle mass |
| $g$ | Gravitational acceleration |
| $A_f$ | Front area |
| $f$ | Rolling resistance coefficient |
| $C_D$ | Air drag coefficient |
| $\rho$ | Air density |
| $\eta_T$ | Transmission efficiency |
| $\eta_{md}$ | Electrical machine efficiency |
| $\eta_r$ | Regenerative braking efficiency |
| $P_V$ | Mechanical power |
| $P_d$ | Vehicle power demand |

*Battery model*
| | |
|---|---|
| $OCV$ | Open-circuit voltage |
| $R_0$ | Ohmic resistor |
| $V_T$ | Terminal voltage |
| $SOC$ | State-of-Charge |
| $Q_{bat}$ | Nominal capacity |
| $\Delta t$ | Sampling time |
| $Q_{loss}$ | Loss in battery capacity |
| $I$ | Current |
| $T$ | Temperature |
| $c_{rate}$ | C-rate |
| $Ah$ | Ampere-hour throughput |
| $R$ | Ideal gas constant |
| $E_a$ | Activation energy |
| $z$ | Battery degradation fitting parameter |
| $a_c$ | Battery degradation fitting parameter |
| $b_c$ | Battery degradation fitting parameter |
| $M$ | Battery degradation pre-exponent factor |
| $SOH$ | State-of-Health |

*Fuel cell model*
| | |
|---|---|
| $\dot{m}$ | Hydrogen mass flow rate |
| $P_{fc}$ | FC output power |
| $a_m$ | Fuel consumption fitting parameter |
| $b_m$ | Fuel consumption fitting parameter |
| $c_m$ | Fuel consumption fitting parameter |
| $V_{FC}$ | FC output voltage |
| $V_{ocv}$ | FC reversible cell voltage |
| $V_{act}$ | FC activation polarization |
| $V_{ohm}$ | FC ohmic voltage losses |
| $V_{con}$ | FC concentration losses |
| $\Delta G$ | Gibbs free energy |
| $F$ | Faraday's constant |
| $T$ | FC temperature |
| $I_{fc}$ | FC current density |
| $R_{fc}$ | FC internal resistance |
| $i_{loss}$ | FC crossover current density |
| $\alpha$ | FC charge transfer coefficients |
| $\beta$ | FC voltage fitting constant |
| $i_0$ | FC exchange current density |
| $i_l$ | FC maximum current density |
| $\Delta V_{load-change}$ | Load change voltage drop rate |
| $\Delta V_{on-off}$ | Start/Stop voltage drop rate |
| $\Delta V_{idling}$ | Idling voltage drop rate |
| $\Delta V_{high-load}$ | High power load voltage drop rate |
| $l_{load-change}$ | Load change degradation |
| $l_{on-off}$ | Start/Stop degradation |
| $l_{idling}$ | Idling degradation |
| $l_{high-load}$ | High power load degradation |
| $\Delta P$ | Power change of the FC system over a sampling time |
| $C_{fc}$ | Cost of a FC stack |
| $V_{drop}^{max}$ | Maximum allowed cumulative voltage drop |
| $P_{fc}^{low}$ | FC low load threshold |
| $P_{fc}^{high}$ | FC high load threshold |

*Cost function*
| | |
|---|---|
| $N$ | total time horizon |
| $M_{fc}$ | total number of FC stacks |
| $M_{bat}$ | total numbers of battery cells |
| $P_d(i)$ | Total power demand at time step $i$ |
| $P_{fc}(i,j)$ | Power of FC stack $j$ at time step $i$ |
| $P_{bat}(i)$ | Battery power at time step $i$ |
| $l_{fc}(i,j)$ | Degradation losses of FC stack $j$ at time step $i$ |
| $l_{bat}(i)$ | Battery degradation losses at time step $i$ |
| $P_{fc}^{min}$ | Minimum FC power |
| $P_{fc}^{max}$ | Maximum FC power |
| $f_{fc-fuel}$ | FC power to hydrogen consumption rate function |
| $f_{bat-I}$ | Battery power to current function |
| $l_{on-off}(i,j)$ | FC load change degradation of FC stack $j$ at time step $i$ |
| $l_{load-change}(i,j)$ | FC start/stop degradation of FC stack $j$ at time step $i$ |
| $l_{high-load}(i,j)$ | FC idling degradation of FC stack $j$ at time step $i$ |
| $l_{idling}(i,j)$ | FC high power load degradation of FC stack $j$ at time step $i$ |
| $I_{bat}(i)$ | Battery current at time step $i$ |
| $SOC_{bat}(i)$ | Battery SOC at time step $i$ |
| $SOC_{bat}^{min}$ | Minimum battery SOC |
| $SOC_{bat}^{max}$ | Maximum battery SOC |
| $SOC_{bat-initial}$ | Battery initial SOC |
| $SOC_{bat-final}^{min}$ | Battery final SOC lower bound |
| $SOC_{bat-final}^{max}$ | Battery final SOC upper bound |

*Fuel cell power to hydrogen mass flow rate*
| | |
|---|---|
| $s_{fc}(i,j)$ | Binary variable indicating whether the FC stack $j$ changed its status (on/off) at time step $i$ |
| $\Delta P_{fc}^{+}(i,j)$ | Absolute value of the power change of FC stack $j$ from time step $i-1$ to $i$ |
| $o_{fc}(i,j)$ | Binary variable indicating the on/off status of the FC stack $j$ at time step $i$ |

*FC degradation loss terms due to high power load and idling reformulation*
| | |
|---|---|
| $h_{fc}(i,j)$ | Binary variable indicating whether FC stack $j$ is in high load condition at time step $i$ |
| $i_{fc}(i,j)$ | Binary variable indicating whether FC stack $j$ is in idling condition at time step $i$ |

*Battery power to current function reformulation*
| | |
|---|---|
| $a_{bat}$ | Battery power to current fitting parameter |
| $b_{bat}$ | Battery power to current fitting parameter |

*Battery degradation term reformulation*
| | |
|---|---|
| $Ah_{EOL}$ | Total Ah throughput of a battery at the end of its life |
| $a_d$ | Battery degradation fitting parameter |
| $b_d$ | Battery degradation fitting parameter |
| $E_{bat}$ | Battery pack energy |
| $C_{bat}$ | Battery price |
| $I_{bat}^{+}(i)$ | Absolute value of the battery power at time step $i$ |

Control (ISC), enabling dynamic control of each stack to operate at its most efficient level.

Rule-based control, popular for real-time control applications, employs predefined rules and thresholds for power distribution between the FC and battery [12]. These rules are typically based on the state of charge (SOC) of the battery, the power demand of the load, and other parameters [13]. A rule-based EMS for FC hybrid trains was developed based on DP results under various conditions [14]. Similarly, [15] generates an optimized rule-based EMS using the genetic algorithm to optimally allocate the power between the FC and the battery system. For multi-stack FC systems, [16] presents three power-splitting algorithms for multiple FC stacks to optimize the efficiency of a multi-FC system. However, the method cannot be directly applied to multi-stack FC-battery hybrid systems and neglects FC system degradation. Macias et al. [17] compared two power allocation strategies for multiple FC stack systems and proposed an adaptive rotary daisy chain strategy to split the power to improve fuel economy, without considering system degradation. Ghaderi et al. [18] combine quadratic programming with rule-based methods to distribute power in a hybrid system, using a very simplified FC degradation model. Similarly, [19] proposed a method using game theory idea to reduce the degradation effects. Rule-based models are unable to account for complex interdependencies within hybrid systems, leading to suboptimal operational outcomes.

Optimization-based control formulates the EMS design as an optimization problem, aiming to find optimal control actions that minimize a specific objective function [20]. Dynamic Programming (DP) and Pontryagin's Minimal Principle (PMP) are among the most common methods because they provide globally optimal solutions. For instance, a DP-based EMS [21] reduces fuel consumption and battery degradation, while a PMP-based EMS focuses on hydrogen consumption reduction [22]. Another study [23] uses an adaptive recursive least square method with PMP to maximize PEMFC power efficiency. However, these global optimization methods often demand significant computational resources, making them ill-suited for real-time applications in complex systems. In contrast, local optimization methods, such as the Equivalent Consumption Minimization Strategy (ECMS) [24], offer faster computation and are suitable for real-time control in dynamic environments. However, local methods may result in suboptimal performance, as they do not guarantee a globally optimal solution.

Learning-based control, such as Neural Networks (NN) and Reinforcement Learning (RL), are also employed to enhance the computational time of optimization-based control methods [25]. These methods may continually adapt and improve control actions through system interactions and feedback. NN can handle complex and non-linear system dynamics, but often requires substantial training data. In [26], eight NNs, trained by selected optimization results, aim to minimize the overall equivalent energy consumption while reducing the high computational cost of optimization. RL utilizes an agent to make decisions through interaction with its environment and feedback using rewards or penalties. A Q-learning-based EMS to minimize battery degradation and energy consumption for hybrid vehicles is proposed in [27]. However, RL may only find a locally optimal or suboptimal solution. Also, RL methods can be computationally expensive as they may require $10^2$ to $10^5$ episodes to converge, depending on the application. Shi et al. [28] employed Q-learning for SOC management and minimizing hydrogen consumption, but RL-based methods still struggle with system transferability and generalization.

Up to date, few works focus on EMS design for multi-stack FC and battery hybrid systems, and the economic operation of such systems that leverage both multi-stack FCs and batteries remains underexplored. Optimal hybrid system performance requires addressing both energy efficiency and the degradation mechanisms of FCs and batteries, making the design of an EMS tailored for FC and battery hybrid systems a challenging task.

In this study, we present a distinct problem reformulation of multi-stack FC and battery hybrid systems that incorporates a more comprehensive consideration. The proposed method not only addresses fuel consumption costs but includes the effects of load changes, on/off switching, idling, and high-power loads on the FC degradation, which significantly contributes to overall degradation compared to other operating modes [29]. The contributions of this study include:

(1) An optimal, health-aware EMS was proposed for managing multiple FC stacks and battery hybrid systems in real time, minimizing both fuel consumption and degradation of FCs and batteries.
(2) A novel reformulation that incorporates multi-stack FC and battery hybrid systems system dynamics and degradation mechanisms into a Mixed-Integer Quadratic Programming (MIQP) structure was proposed, which is suitable for real-time applications.
(3) The benefits of ISC were presented, showing that independently controlling FC stacks, rather than as a collective stack, improves system efficiency and longevity.

## 2. Model

This study employs an electric hybrid city bus as an application for the FC and battery hybrid system. The first subsection details the vehicle dynamics and powertrain configuration. The other two subsections detail the battery and FC models.

### 2.1. Vehicle model and configurations

Fig. 1 illustrates the structure of the hybrid energy storage system. The electric motor propels the rear wheels. Positioned between the electric motor and the two energy storage systems is a DC/AC converter. To balance the voltage between the battery pack and FC stacks, an AC/DC converter and DC/DC converters are utilized [30]. In the hybrid system, there is one battery pack and multiple FC stacks.

The vehicle dynamics model formulation is widely used in the field [31], [32], [33] and can be expressed as follows,

$$P_V = m\,g\,f\,v\,cos(\alpha) + 0.5\,C_D\,A_f\,\rho\,v^3 \\ + m\,v\,\frac{dv}{dt} + m\,g\,v\,sin(\alpha) \quad (1)$$

$$P_d = \begin{cases} P_V > 0: & P_V\,\eta_T\,\eta_{md} \\ P_V \leq 0: & \dfrac{P_V}{\eta_r} \end{cases} \quad (2)$$

Here, $P_V$ denotes the mechanical power required at the wheel to counteract road friction, air drag, gravity, etc. The electric power demand ($P_d$) is determined by transmission efficiency ($\eta_T$), electrical machine efficiency ($\eta_{md}$), and regenerative braking efficiency ($\eta_r$). Despite the simplified nature of the adopted vehicle dynamics model, it captures key physical factors (e.g., aerodynamic drag, inertial forces, rolling resistance, and grade resistance), ensuring the generated load profiles closely represent actual vehicle operations under variable and demanding driving conditions. Other parameters are in Table 1 [34].

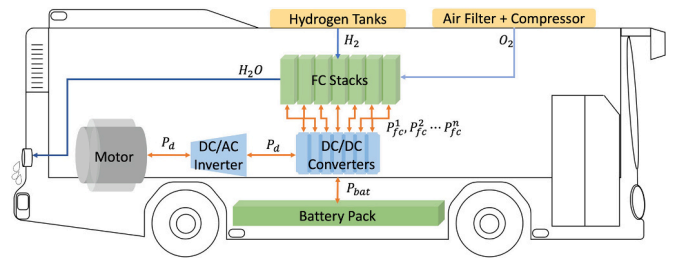

**Fig. 1.** Architecture of the multi-stack FC-battery hybrid systems bus propulsion system.

## 2.2. Battery model

(1) Battery Electrical Model

An equivalent circuit model (ECM), also known as the Rint model, is utilized to simulate the electrical characteristics of a battery. The ECM comprises an open-circuit voltage (OCV) and an ohmic resistor ($R_0$). The terminal voltage ($V_T$) is the output of ECM. The values for the OCV and $R_0$ are dependent on the SOC (in %), temperature, and both discharge and charge currents.

The dynamics of a battery can be formulated as follows,

$$SOC(i) = SOC(i-1) + \frac{\Delta t\, I}{3600\, Q_{bat}} \cdot (100\%) \tag{3}$$

$$V_T = OCV + R_0 I \tag{4}$$

where $I$ (in A) is the current, $\Delta t$ (in s) is sampling time, and $Q_{bat}$ (in Ah) is the current nominal capacity.

(2) Battery Degradation Model

An empirical battery degradation model was adopted, which is widely used to simulate Li-ion battery degradation behaviors [35]. This model describes the loss in battery capacity ($Q_{loss}$, in %) as a function of temperature ($T$, in K), C-rate ($c_{rate}$), and Ampere-hour throughput ($Ah$),

$$Q_{loss} = M(c_{rate})\, exp\left(\frac{-E_a}{RT}\right)(Ah)^z \tag{5}$$

$$Ea = a_c + b_C\, C_{rate} \tag{6}$$

where R signifies the ideal gas constant, while $E_a$ is the activation energy. The power law factor is represented by $z$ and set to 0.55. The two fitting parameters $a_c$ and $b_c$ are 31,700 and 370.3. $M$, is a pre-exponent factor that diminishes as the C-rate increases, as demonstrated in Table 2 [36].

The term $Ah_{EOL}$ is used to refer to the Ampere-hour throughput value at which the battery reaches its End-of-Life (EOL), which is defined as the point when the battery has lost 20 % of its initial capacity. As shown in Eq. (5), the cycles until EOL depends on the operating conditions, such as C-rate and temperature.

Additionally, the battery's State of Health (SOH) is defined as the ratio of the battery's current capacity to its initial capacity. The SOH value can be determined using the degradation model employed in this study, which is described as follows,

$$SOH = \frac{Q_{initial} - Q_{loss}}{Q_{initial}} \times 100\% = \frac{Q_{bat}}{Q_{initial}} \times 100\% \tag{7}$$

where $Q_{initial}$ represents the initial capacity of a battery, $Q_{loss}$ indicates the capacity loss over time due to degradation, and $Q_{bat}$ is the current nominal capacity.

**Table 1**
Parameters and description of the electric city bus.

| Parameter | Description | Value |
|---|---|---|
| $m$ | Vehicle mass | 13500 kg |
| $g$ | Gravitational acceleration | 9.8 m/s$^2$ |
| $A_f$ | Front area | 7.5 m$^2$ |
| $f$ | Rolling resistance coefficient | 0.018 |
| $C_D$ | Air drag coefficient | 0.7 |
| $\rho$ | Air density | 1.29 kg/m$^3$ |
| $\eta_T$ | Transmission efficiency | 90% |
| $\eta_{md}$ | Electrical machine efficiency | 85% |
| $\eta_r$ | Regenerative braking efficiency | 50% |

**Table 2**
Pre-exponential factor as a function of the C-rate.

| C-rate | 0.5 | 2 | 6 | 10 |
|---|---|---|---|---|
| M | 31,630 | 21,681 | 12,934 | 15,512 |

## 2.3. FC Model

Proton Exchange Membrane (PEM) FCs are favored in hybrid energy systems due to their high energy efficiency, fast startup, response time, low operating temperature, compact design, environmental friendliness, quiet operation, and scalability [37]. As such, all the FCs utilized in this study are assumed to be of the PEM type.

(1) FC Power to Hydrogen Mass Flow Rate Curve

Fig. 2 presents the consumption and efficiency curve of an FC stack (consisting of 500 individual FCs, each possessing an effective area of 280 cm$^2$) derived from Advanced Vehicle Simulator (ADVISOR, a scaled version of FC_ANL50H2).

The relationship between hydrogen mass flow rate and the power of a FC per unit effective catalyst surface area is illustrated in Fig. 3. A quadratic function has been used to approximate the mapping of power output to hydrogen consumption, as depicted below,

$$\dot{m} = a_m P_{fc}^2 + b_m P_{fc} + c_m \tag{8}$$

In this equation, $\dot{m}$ signifies the hydrogen mass flow rate, $P_{fc}$ is the output power of a FC per unit effective catalyst surface area, with $a_m$, $b_m$, and $c_m$ being fitting parameters.

(2) FC Degradation Model

The voltage of a FC is determined by the polarization equation as shown below:

$$V = V_{ocv} - V_{act} - V_{ohm} - V_{con} \tag{9}$$

where $V_{ocv}$, $V_{act}$, $V_{ohm}$, and $V_{con}$ refer to reversible voltage, activation polarization, ohmic voltage losses, and concentration polarization, respectively. The respective equations for these terms are,

$$V_{ocv} = -\frac{\Delta G}{2F} \tag{10}$$

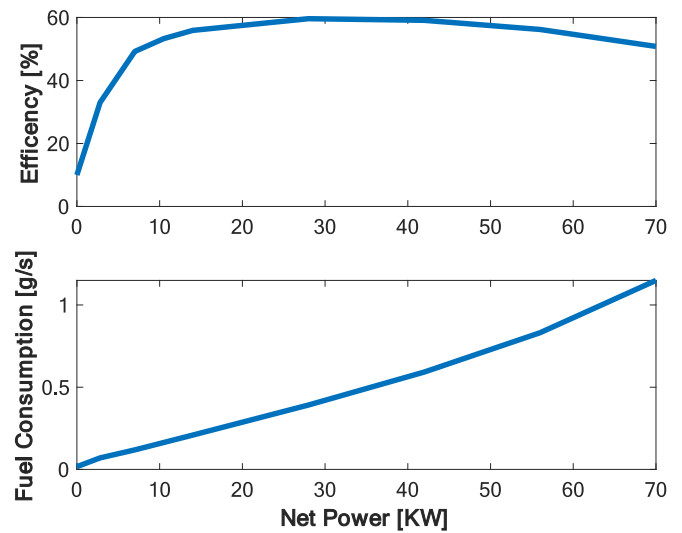

Fig. 2. FC efficiency and fuel consumption curves.

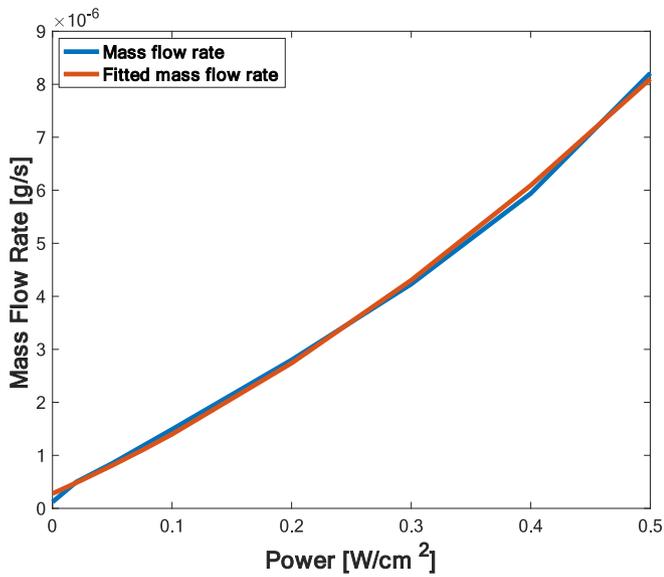

**Fig. 3.** FC power to hydrogen mass flow rate curve.

$$V_{act} = \alpha \, T \ln\left(\frac{I_{fc} + i_{loss}}{i_0}\right) \quad (11)$$

$$V_{ohm} = R_o \, I_{fc} \quad (12)$$

$$V_{con} = -\beta \ln\left(1 - \frac{I_{fc}}{i_l}\right) \quad (14)$$

where $\Delta G$ is the Gibbs free energy, $F$ is Faraday's constant, $T$ is temperature, and $I_{fc}$ is current density. $R_o$ is FC internal resistance, $\alpha$ is charge transfer coefficients, $i_{loss}$ is crossover current density, $i_0$ is exchange current density, $\beta$ is a fitted constant, and $i_l$ is the maximum current density.

Certain operations, such as altering load, toggling the start-stop condition, idling, and handling high-power loads, significantly contribute to FC degradation [38], [39]. These operation conditions can lead to several detrimental outcomes, including corrosion, thermal stress/cycling, mechanical stress, drying/flooding, and impurities within FC systems [40], [41], [42]. As depicted in Fig. 4, these conditions subsequently cause degradation to the FCs' components, such as membrane, catalyst, gas diffusion layer, and bipolar elements [43], [44]. These degraded components result in a decrease in charge transfer coefficient, maximum current density, an increase in internal current density, exchange current density, ohmic resistance, and the concentration loss constant [45]. The changes in $R_o$, $\alpha$, $\beta$, $i_{loss}$, and $i_0$, contribute to an increase in the voltage loss terms in the polarization equation, leading to power fade in the FC system.

In Table 3, the voltage drop rates for the above-mentioned four conditions are referred to as $\Delta V_{load-change}$, $\Delta V_{on-off}$, $\Delta V_{idling}$, and $\Delta V_{high-load}$. If cumulative voltage degradation surpasses 10 % of the rated voltage of a new FC, the FC is regarded as at its end-of-life and needs replacement due to its diminished power efficiency [21].

The increased component costs associated with the degradation of the FC system under four different operating conditions, $l_{load-change}$, $l_{on-off}$, $l_{idling}$, and $l_{high-load}$, can be calculated using the prescribed voltage degradation rates as follows,

Load change:

$$l_{load-change} = \frac{|\Delta P| \Delta V_{load-change} \, C_{fc}}{V_{drop}^{max}} \quad (14)$$

On/off:

$$l_{on-off} = \begin{cases} \dfrac{\Delta V_{on-off} \, C_{fc}}{V_{drop}^{max}} & \text{if on/off triggered} \\ 0 & \text{otherwise} \end{cases} \quad (15)$$

Idling:

$$l_{idling} = \begin{cases} \dfrac{\Delta t \Delta V_{idling} \, C_{fc}}{3600 \, V_{drop}^{max}} & \text{if } 0 < P_{fc} \leq P_{fc}^{low} \\ 0 & \text{if } P_{fc} > P_{fc}^{low} \end{cases} \quad (16)$$

High power load:

$$l_{high-load} = \begin{cases} \dfrac{\Delta t \Delta V_{high-load} \, C_{fc}}{3600 \, V_{drop}^{max}} & \text{if } P_{fc} \geq P_{fc}^{high} \\ 0 & \text{if } P_{fc} < P_{fc}^{high} \end{cases} \quad (17)$$

where $\Delta P$ (in kW) is the power change of the FC system over a sampling time, $C_{fc}$ (in \$) represents the total cost of FC stacks, $V_{drop}^{max}$ (in μV) indicates the maximum allowed cumulative voltage drop of FC, and $\Delta t$ (in second) represents the sampling time of the system. The thresholds for low and high power denoted as $P_{fc}^{low}$ and $P_{fc}^{high}$, are designated as 0.2 and 0.8 of the maximum power supply capability of FC system, respectively [47]. These losses will be used in the optimization objective to disincentivize operations that cause degradation.

## 3. Methodology

In this section, the first subsection focuses on the cost function of the hybrid system. The subsequent part shows reformulations of constraints within the cost function. The constraints and system dynamics are explicitly restructured into a convex quadratic form to facilitate efficient optimization using MIQP. This approach significantly reduces computational complexity while maintaining high-fidelity system representation. In the final subsection, the selection of the solver for the MIQP is presented. Throughout this section, blue variables are optimization variables. Orange variables are the terms in the objective function.

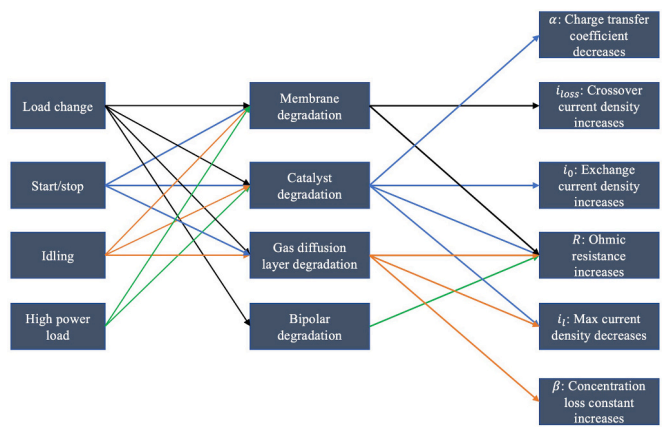

**Fig. 4.** FC degradations caused by four operational conditions.

**Table 3**
Voltage degradation rates [46].

| Operations: | Symbols: | Drop rate: |
| --- | --- | --- |
| Load change | $\Delta V_{load-change}$ | 1.79 μV/kW |
| Start/Stop | $\Delta V_{on-off}$ | 13.79 μV |
| Idling | $\Delta V_{idling}$ | 8.66 μV/h |
| High power load | $\Delta V_{high-load}$ | 10 μV/h |

## 3.1. Cost function of the hybrid system

To minimize the operational cost over a specified time horizon and power demand profile, determining the optimal sequential actions for each FC stack and the battery pack in the hybrid system becomes imperative. This goal involves minimizing the degradation losses of both the FC stacks and the battery pack, along with the fuel consumption cost of the FCs. Thus, we can formulate the goal as an optimization problem, represented by the following equation,

$$J = \sum_{i=1}^{N} \sum_{j=1}^{M_{fc}} \dot{m}(i,j) \Delta t \, C_{H_2} + l_{fc}(i,j) + M_{bat} \, l_{bat}(i) \tag{18}$$

where $i$ denotes the current time step within the total time horizon $N$, while $j$ corresponds to a specific FC stack, with $M_{fc}$ being the total number of FC stacks in the system. The fuel consumption cost is calculated based on the hydrogen mass fuel rate $\dot{m}(i,j)$ (in $kg/s$) of a given FC stack $j$ at time $i$, the sampling time $\Delta t$ (in seconds), and the hydrogen price $C_{H_2}$ (in $\$/kg$). The loss term, $l_{fc}(i,j)$ represents the degradation losses (in $\$$) incurred by the FC stack $j$ at time step $i$. $M_{bat}$ is the total numbers of cells in a battery pack. $M_{bat} \times l_{bat}(i)$ represents total degradation losses (in $\$$) for the battery pack.

The cost function is subject to several constraints, including power constraints, FC dynamic constraints, battery dynamic constraints, and initial and terminal constraints. For all time steps $i$, the constraints are described below,

Power constraint:

$$P_d(i) = P_{bat}(i) + \sum_{j=1}^{M_{fc}} P_{fc}(i,j) \# \tag{19}$$

$$P_{fc}^{min} \leq P_{fc}(i,j) \leq P_{fc}^{max} \tag{20}$$

FC fuel consumption and degradation constraints:

$$\dot{m}(i,j) = f_{fc-fuel}(P_{fc}(i,j)) \tag{21}$$

$$l_{fc}(i,j) = l_{on-off}(i,j) + l_{load-change}(i,j) \\ + l_{high-load}(i,j) + l_{idling}(i,j) \tag{22}$$

Battery dynamic and degradation constraints:

$$I_{bat}(i) = f_{bat-I}\left(\frac{P_{bat}(i)}{M_{bat}}, SOC_{bat}(i)\right) \tag{23}$$

$$I_{bat}^{min} \leq I_{bat}(i) \leq I_{bat}^{max} \tag{26}$$

$$SOC_{bat}(i+1) = SOC_{bat}(i) + \frac{100\% \Delta t \, I_{bat}(i)}{3600 \, Q_{bat}} \tag{25}$$

$$SOC_{bat}^{min} \leq SOC_{bat}(i) \leq SOC_{bat}^{max} \tag{26}$$

$$l_{bat}(i) = f_{bat-loss}(I_{bat}(i)) \tag{27}$$

Initial and terminal constrains:

$$SOC_{bat}(1) = SOC_{bat-initial} \tag{28}$$

$$SOC_{bat-final}^{min} \leq SOC_{bat}(N) \leq SOC_{bat-final}^{max} \tag{29}$$

$f_{fc-fuel}(P_{fc}(i,j))$ refers to Eq. (8) and is used to calculate the hydrogen consumption rate, $\dot{m}$, based on the FC power and the FC power to mass flow rate curve, as shown in Fig. 3. $l_{fc}(i,j)$ links the total FC degradation of stack $j$ at time step $i$ to individual degradation terms. $l_{on-off}(i,j)$, $l_{load-change}(i,j)$, $l_{high-load}(i,j)$, and $l_{idling}(i,j)$ refer to Eq. (14–17) and represent degradation of the FC system under four different operating conditions.

It is noted that terms in the objective function and constraints, including $f_{fc-fuel}(P_{fc}(i,j))$, $f_{bat-loss}(I_{bat}(i))$, $l_{on-off}(i,j)$, $l_{load-change}(i,j)$, $l_{high-load}(i,j)$, $l_{idling}(i,j)$, and $f_{bat-I}\left(\frac{P_{bat}(i)}{M_{bat}}, SOC_{bat}(i)\right)$ are highly nonlinear.

The following subsections will delve into the reformulation of these highly nonlinear terms to into a MIQP optimization problem. The reformulation process aims to transform these nonlinear terms into linear or quadratic forms, facilitating their incorporation into the optimization framework.

## 3.2. FC power to hydrogen mass flow rate

To accurately capture the on-off mechanism of the FC stack, a binary control variable, $o_{fc}(i,j)$, is introduced to represent the "on" condition of the FC stack $j$ at time step $i$. The following constraints ensure that $o_{fc}(i,j)$ is set to zero when the power output of the FC stack, $P_{fc}(i,j)$, is lower than $P_{fc}^{min}$, and it is set to one otherwise,

$$o_{fc}(i,j) \, P_{fc}^{min} \leq P_{fc}(i,j) \tag{30}$$

$$P_{fc}(i,j) \leq o_{fc}(i,j) \, P_{fc}^{max} \tag{31}$$

Then, $f_{fc-fuel}(P_{fc}(i,j))$ can be reformulated as a quadratic function and included in the objective function,

$$\dot{m}(i,j) = a_{fc} \, P_{fc}(i,j)^2 + b_{fc} \, P_{fc}(i,j) \\ + c_{fc} \, o_{fc}(i,j) \# \tag{32}$$

where $a_{fc}$, $b_{fc}$, and $c_{fc}$ are fitted parameters that capture the characteristics of the power-to-hydrogen mass flow for the FC stack. This reformulation ensures that the hydrogen mass flow rate is forced to zero when the FC stack is turned off, allowing for accurate of the FC stack's behavior and power output and the on-off status representation.

## 3.3. FC loss due to on-off switch and load changes

Calculating the FC loss terms requires determining whether an on-off switch has occurred and expressing the absolute value of the power change. To capture these aspects, we introduce two variables for each time step $i$ and FC stack $j$: the binary variable $s_{fc}(i,j)$ and the positive variable $\Delta P_{fc}^+(i,j)$.

The binary control variable $s_{fc}(i,j)$ is set to one if there is a change from $o_{fc}(i-1,j)$ to $o_{fc}(i,j)$. Otherwise, it is forced to zero by the cost function. The following constraints represent this relationship,

$$s_{fc}(i,j) \geq o_{fc}(i,j) - o_{fc}(i-1,j) \tag{33}$$

$$s_{fc}(i,j) \geq o_{fc}(i-1,j) - o_{fc}(i,j) \tag{34}$$

For example, if the status of FC stack j changes from 'off' to 'on' at time step $i$, then in the above two constraints, $o_{fc}(i,j) - o_{fc}(i-1,j)$ would be 1 and $o_{fc}(i-1,j) - o_{fc}(i,j)$ would be $-1$, so $s_{fc}(i,j)$ will be constrained to 1.

The variable $\Delta P_{fc}^+(i,j)$ is constrained to represent the absolute value of the power change, $|P_{fc}(i,j) - P_{fc}(i-1,j)|$, using the following equations,

$$\Delta P_{fc}^+(i,j) \geq P_{fc}(i-1,j) - P_{fc}(i,j) \tag{35}$$

$$\Delta P_{fc}^+(i,j) \geq P_{fc}(i,j) - P_{fc}(i-1,j) \tag{36}$$

With the above two constraints and the cost function, the value of $\Delta P_{fc}^+(i,j)$ will always take the larger values of $P_{fc}(i-1,j) - P_{fc}(i,j)$ and $P_{fc}(i,j) - P_{fc}(i-1,j)$ to capture the absolute power change of FC stack $j$ from time step $i-1$ to $i$.

With these variables in place, we can rewrite the Eq. (14) and Eq. (15) and express the FC degradation loss terms $l_{load-change}(i,j)$ and $l_{on-off}(i,j)$ as follows,

$$l_{load-change}(i,j) = \frac{\Delta V_{load-change} \, C_{fc}}{V_{drop}^{max}} \Delta P_{fc}^+(i,j) \tag{37}$$

$$l_{on-off}(i,j) = \frac{\Delta V_{on-off}\, C_{fc}}{V_{drop}^{max}} s_{fc}(i,j) \tag{38}$$

where $\Delta V_{on-off}$ and $\Delta V_{load-change}$ are voltage degradation rates due to on-off switching and load changing, as shown in Table 3. $C_{fc}$ is the price of the FC, and $V_{drop}^{max}$ is the maximum allowed cumulative voltage drop due to degradation. These formulations ensure that the loss terms accurately capture the effects of on/off switching and load changing on the FC stack.

### 3.4. FC degradation loss terms due to high power load and idling reformulation

It is necessary to identify whether an FC stack is operating under high power load or idling conditions when calculating the FC degradation loss terms. We introduce two variables for each time step $i$ and FC stack $j$ to capture these aspects: the binary variables $h_{fc}(i,j)$ and $i_{fc}(i,j)$.

To enforce the binary variable $h_{fc}(i,j)$ to indicate whether the power of FC stack $j$ is higher than $P_{fc}^{high}$ at time step $i$, the following constraints are utilized,

$$h_{fc}(i,j) > \frac{P_{fc}(i,j) - P_{fc}^{high}}{P_{fc}^{max} - P_{fc}^{high}} \tag{39}$$

$$h_{fc}(i,j) \leq \frac{P_{fc}(i,j) - P_{fc}^{min}}{P_{fc}^{high} - P_{fc}^{min}} \tag{40}$$

For example, if $P_{fc}(i,j)$ is higher than $P_{fc}^{high}$, the term $\frac{P_{fc}(i,j) - P_{fc}^{high}}{P_{fc}^{max} - P_{fc}^{high}}$ in Eq. (39) would yield a value larger than 0 and smaller than 1, and the term $\frac{P_{fc}(i,j) - P_{fc}^{min}}{P_{fc}^{high} - P_{fc}^{min}}$ in Eq. (40) would yield a greater than 1. Since $h_{fc}(i,j)$ is a binary variable, the two above constraints would force $h_{fc}(i,j)$ to be 1 in the case.

On the other hand, if $P_{fc}(i,j)$ is lower than $P_{fc}^{high}$, the term in Eq. (39) would yield a value less than 0, and the term in Eq. (40) would yield a value between 0 and 1, forcing $h_{fc}(i,j)$ to be 0.

Similarly, to ensure that the binary variable $i_{fc}(i,j)$ represents whether the power of FC stack $j$ is lower than $P_{fc}^{low}$ at time step $i$, the following constraints are applied,

$$i_{fc}(i,j) > \frac{P_{fc}^{low} - P_{fc}(i,j)}{P_{fc}^{low} - P_{fc}^{min}} \tag{41}$$

$$i_{fc}(i,j) \leq \frac{P_{fc}^{max} - P_{fc}(i,j)}{P_{fc}^{max} - P_{fc}^{low}} \tag{42}$$

By incorporating these binary control variables, the loss terms $l_{idling}(i,j)$ and $l_{high-load}(i,j)$ and Eq. (16) and Eq. (17) can be expressed as follows,

$$l_{idling}(i,j) = \frac{\Delta t \Delta V_{idling}\, C_{fc}\, (i_{fc}(i,j) + o_{fc}(i,j) - 1)}{3600\, V_{drop}^{max}} \tag{43}$$

$$l_{high-load}(i,j) = \frac{\Delta t \Delta V_{high-load}\, C_{fc}\, h_{fc}(i,j)}{3600\, V_{drop}^{max}} \tag{44}$$

Because binary variable $o_{fc}(i,j)$ is set to 1 only if the FC stack $j$ is "on" at the time step $i$, the inclusion of $o_{fc}(i,j)$ in the Eq. (43) ensures that the $l_{idling}(i,j)$ term is equal to zero when the FC stack $j$ is turned off.

### 3.5. Battery power to current function reformulation

The battery power to current function, given the output power of the ECM, can be represented by the following nonlinear equation,

$$I_{bat} = \frac{V_{ocv}(SOC) - \sqrt{V_{ocv}(SOC)^2 - 4R(SOC) P_{bat}}}{2R(SOC)}\# \tag{45}$$

Although this function is nonlinear, a closer examination of the relationship between the output power, SOC, and battery current reveals that the surface is not excessively nonlinear, as depicted in Fig. 5.

Based on this observation, a linear regression approach is used to approximate the battery power to current relationship, $f_{bat-I}\left(\frac{P_{bat}(i)}{M_{bat}}, SOC_{bat}(i)\right)$ described in Eq. (23). This relationship is fitted to the linear equation,

$$I_{bat}(i) = a_{bat}\frac{P_{bat}(i)}{M_{bat}} + b_{bat}\, SOC_{bat}(i) \tag{46}$$

where $a_{bat}$ and $b_{bat}$ are fitted parameters. The R-squared value of 0.9873 indicates a strong correlation between the fitted linear model and the actual battery dynamics. Therefore, this linear approximation is sufficient.

### 3.6. Battery degradation term reformulation

For battery degradations, a lower C-rate charge and discharge operations leads to a higher $Ah_{EOL}$, indicating a longer battery life. The $Ah_{EOL}$, corresponding to a 20 % capacity loss, can be determined for each C-rate, representing the EOL condition (SOH = 80 %) of the battery. Fig. 6 demonstrates the relationship between the C-rate ($c_{rate}$) and $\frac{1}{Ah_{EOL}}$. It is notable that this relationship exhibits a nearly linear trend.

This allows for the linear fitting to approximate the relationship between the C-rate and $\frac{1}{Ah_{EOL}}$ as

$$\frac{1}{Ah_{EOL}} = a_d\, c_{rate} + b_d \tag{47}$$

where $a_d$ and $b_d$ are linear fitting parameters.

To calculate the battery degradation loss in a sampling period, we divide the change in Ah throughput ($\Delta Ah$) in the sampling time by $Ah_{EOL}$, and then multiply it by the battery pack energy ($E_{bat}$, in kWh) and the battery price ($C_{bat}$, in $/kWh). This can be expressed as,

$$l_{bat} = \frac{\Delta Ah}{2\, Ah_{EOL}}\, E_{bat}\, C_{bat} \tag{48}$$

Substituting current into the above equation, $f_{bat-loss}(I_{bat}(i))$ can be reformulated as a quadratic form with the given current,

$$I_{bat}^+(i) >= -I_{bat}(i) \tag{49}$$

$$I_{bat}^+(i) >= I_{bat}(i) \tag{50}$$

$$l_{bat}(i) = \left( a_d \frac{I_{bat}^+(i)^2}{Q_{bat}} + b_d I_{bat}^+(i) \right) \frac{\Delta t}{7200} E_{bat} C_{bat} \tag{51}$$

where $I_{bat}^+(i)$ represents the absolute value of the battery current. This formulation enables the integration of battery degradation cost into the objective function in a quadratic form.

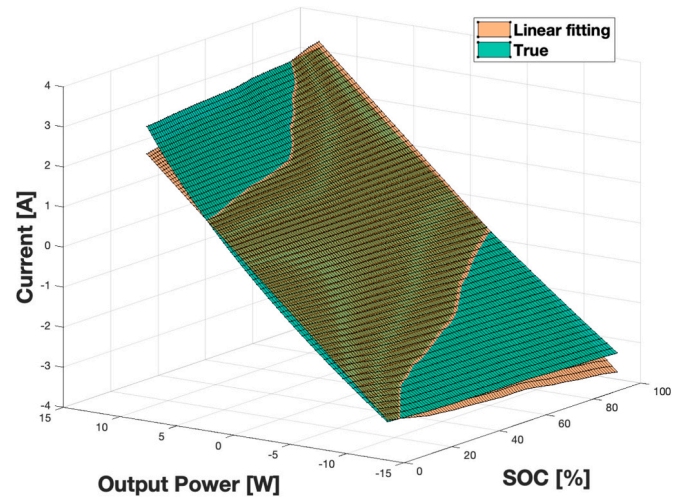

**Fig. 5.** The current of a 3.2 Ah battery cell across different SOC and output power ranges.

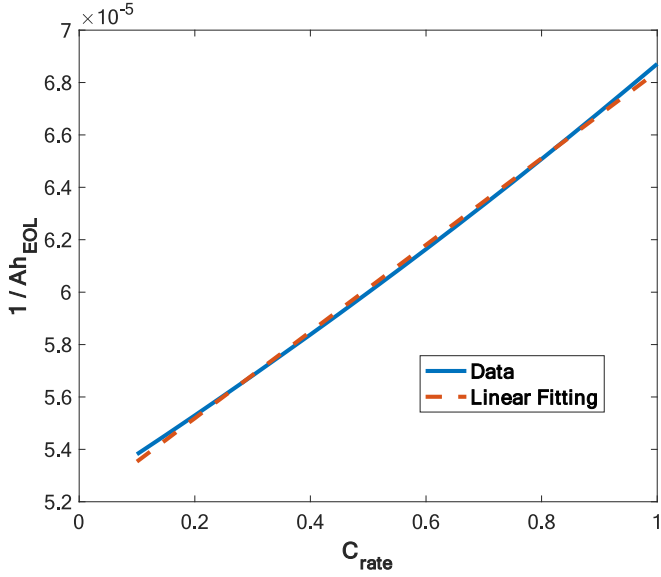

**Fig. 6.** *C-rate* vs. $\frac{1}{Ah_{EOL}}$.

### 3.7. MIQP solver selection

The solver selection is critical for the online operation of the EMS because it must provide rapid and efficient optimization suitable for real-time applications. We employed Gurobi [48], a well-regarded commercial software package, as our solver.

## 4. Performance and discussion

In this section, we first discuss the setup of the test environment, followed by an examination of two distinct control methods:

(1) **Collective Stack Control (CSC):** All FC stacks are treated as a single unit, with identical control actions applied simultaneously.
(2) **Individual stack control (ISC):** Each FC stack is controlled independently, with tailored actions to enhance fuel efficiency and minimize degradation

### 4.1. System setup

In the simulation, the proposed method is evaluated using the Chinese Bus Driving Cycle (CBDC) profile in Fig. 7. In the lower subplot, the power demand profile is shown, which is generated from the CBDC speed profile in the upper subplot.

The total optimization horizon is set to 600 s, with a sampling time of 1 s. The online optimization is performed every 60 s using a block MPC approach. This means the system calculates a block of control actions $u(t), u(t+1), \ldots, u(t+59)$ every 60 s [31]. The initial values for the next

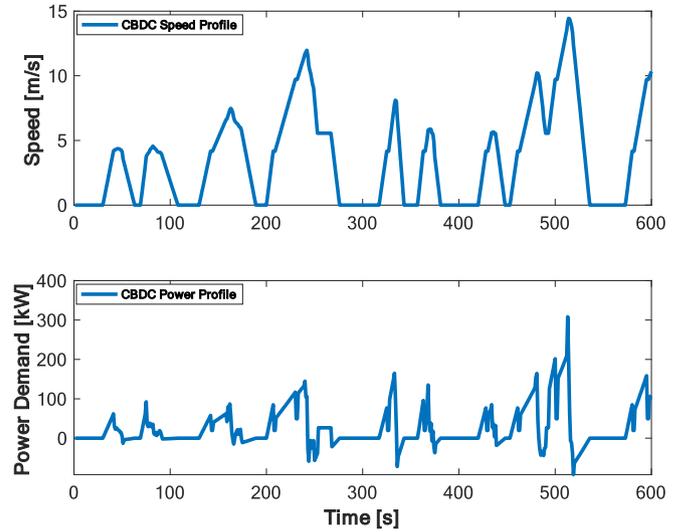

**Fig. 7.** CBDC speed and power profile.

optimization are updated based on measurements at the block frequency.

In Fig. 8, the hybrid city bus updates its states, such as SOC and control actions, every 60 s. It allows the controller/online optimization system to generate optimal reference control actions with a 600-s prediction horizon, balancing short-term responsiveness and computational efficiency.

In the hybrid city bus under study, the powertrain includes eight FC stacks and one battery pack. Referring to Eq. (18), $N$ is 600, $M_{fc}$ is 8, and $\Delta t$ is 1. Additionally, the hydrogen price $C_{H_2}$ is set to be \$4/kg [49]. The FC system consists of eight stacks, each comprising 500 individual FCs connected in series, which allows the system to generate a maximum output of 70 kW per stack. Each stack has an effective area of 280 cm$^2$. These stacks are connected in parallel to meet required total power. Refering to Eq. (20), $P_{fc}^{min}$ and $P_{fc}^{max}$ are 20 % and 80 % of the maximum power supply capability of the FC system, which correspond to 14 and 56 kW, respectively.

The battery pack consists of 7594 individual cells, each with a capacity of 3.2 Ah. The cells are connected in a combination of serial and parallel arrangements to meet the voltage and current requirements of the vehicle, providing a total energy capacity of 90 kWh. The maximum charge and discharge C-rate is 1.2C, allowing the battery pack to operate efficiently in conjunction with the FC system. The cell number in a battery pack, $M_{bat}$, is 7594. In Eq. (23) and (24), $I_{bat}^{min}$, $I_{bat}^{max}$, and $Q_{bat}$ are −3.84 A, 3.84 A and 11,520 As, respectively.

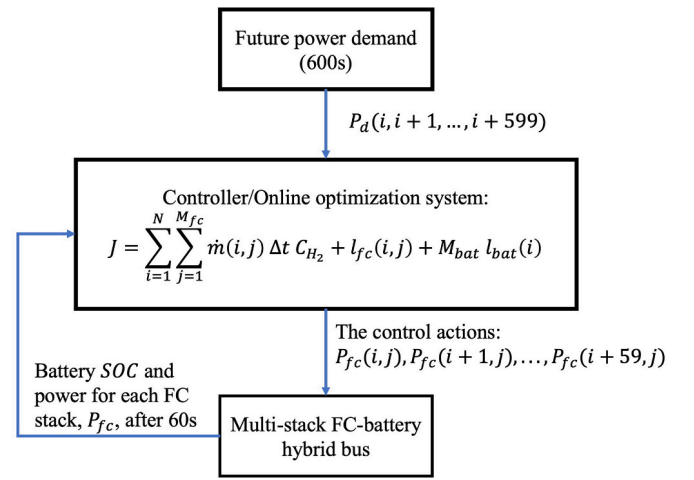

**Fig. 8.** Frame and flow chart of the proposed system.

We set the battery initial SOC to 50 %. To ensure that the battery retains the capability to provide and absorb energy in the future, we impose a terminal constraint that limits the battery's SOC to between 47 % and 53 % at the end of the optimization period. Therefore, $SOC_{bat-initial}$, $SOC_{bat-final}^{min}$, and $SOC_{bat-final}^{max}$ in the Eq. (28) and (29) are 50 %, 47 % and 53 %, respectively.

Additionally, we set the costs of the FC and battery at $960/kW [50] and $178.41/kWh [51], respectively, for $C_{fc}$ and $C_{bat}$ in Eq. (37), (38), (43), (44) and (51).

**Table 4**
System characteristics.

| Parameters | Values |
| --- | --- |
| $N$ | 600 s |
| $M_{fc}$ | 8 |
| $M_{bat}$ | 7594 |
| $\Delta t$ | 1 s |
| $C_{H_2}$ | $4/kg |
| $P_{fc}^{min}$ | 14 kW |
| $P_{fc}^{max}$ | 56 kW |
| $SOC_{bat-initial}$ | 50% |
| $SOC_{bat-final}^{min}$ | 47 % |
| $SOC_{bat-final}^{max}$ | 53 % |
| $C_{fc}$ | $960/kW |
| $C_{bat}$ | $178.41/kWh |

While typical electric buses provide maximum power in the range of 250 kW to 660 kW [11], the maximum power in the test power profile is about 310 kW, which is relatively low compared to the capacity of the electric bus system used in this study. In this study, the maximum power the electric city bus system can provide is about 650 kW. The power profile was selected to highlight the importance of using the ISC rather than the CSC method, particularly in cases of low power demand.

*4.2. Benchmark approach: DP method*

DP was selected as the benchmark in this study due to its ability to compute globally optimal control policies under known dynamics. Since the objective of our proposed MIQP method is to approximate DP-level optimality while improving scalability and execution time, we chose DP as the sole benchmark for focused and meaningful comparison. However, due to the high computational cost of DP, all FC stacks are considered as a single large stack, and are assigned the same control actions using the CSC way.

DP involves discretizing the state and control space into finite points and solving Bellman's equation in a backward time manner iteratively [20]. For each discretized state $x$, including battery SOC and previous FC output power, and for each time step $i$, the cost of all potential control actions is calculated. In this study, the grid sizes of the DP method for the SOC and FC output power are set to 0.02 % and 5 kW, respectively. The control, $u_i$, that minimizes the cost is chosen. The cost not only includes the immediate cost at the current time step, but also the future cost obtained from the next time step's value function which is calculated in the previous iteration as,

$$V(\mathbf{x}_i, i) = \min_{u_i}\{C_i(\mathbf{x}_i, u_i) + V(\mathbf{x}_{i+1}, i+1)\} \quad (52)$$

The total cost resulting from this policy can be compared to the total cost obtained from the proposed algorithm.

*4.3. Comparative analysis of MIQP and DP methods with collective stack control*

Fig. 9 illustrates the power split profiles between the FC and the battery using the DP and MIQP methods, respectively. The power split profiles derived from both the DP and MIQP methods are similar. The battery buffers transient power spikes and provides additional power during high-load conditions. This alleviates the strain on the FCs and reduces the occurrence of load changes and high-load conditions that lead to FC degradation. In low-load conditions, the battery effectively acts as an energy storage device, allowing the FCs to avoid idling and start-stop conditions by absorbing excess power through charging. Overall, the battery helps regulate and smooth out the power demand, enabling the FCs to operate within their optimal power range and leading to an extended lifespan for the FCs.

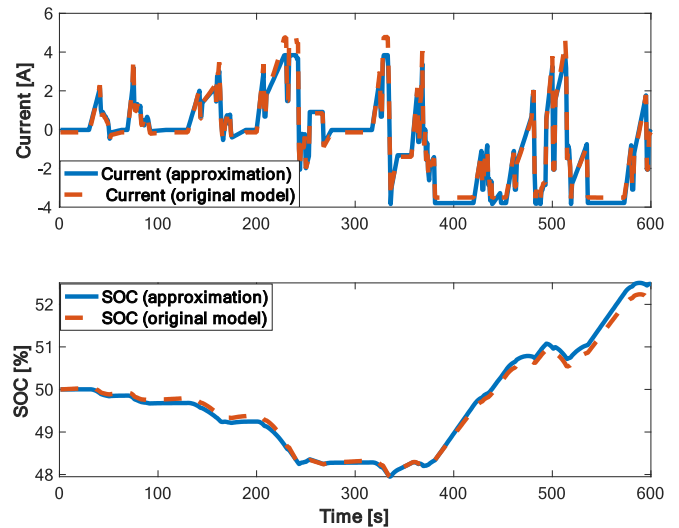

**Fig. 10.** Comparison of battery current and SOC: approximation vs original Model.

It's crucial that the approximation of battery power to current and the reformulation of the battery degradation term serve as foundational elements for the application of MIQP. Fig. 11 and Fig. 12 illustrate a comparison of battery current, SOC, and degradation between the approximation and the original model. This approximation accurately shows the relationship between power and current under battery's

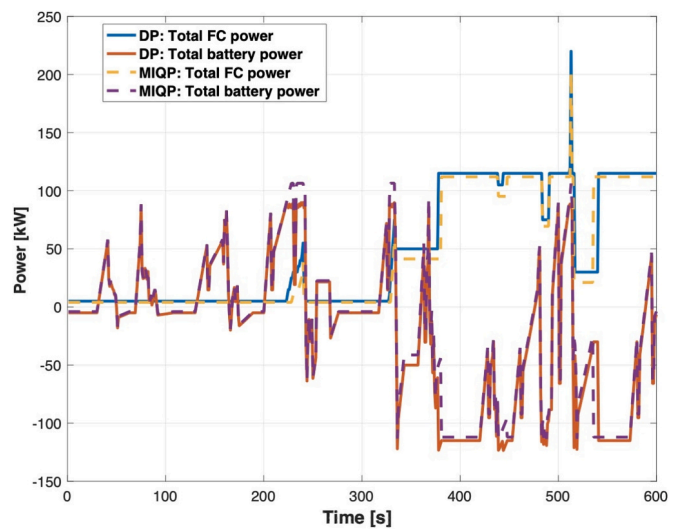

**Fig. 9.** Power split profiles from DP and MIQP results using the Collective Stack Control method.

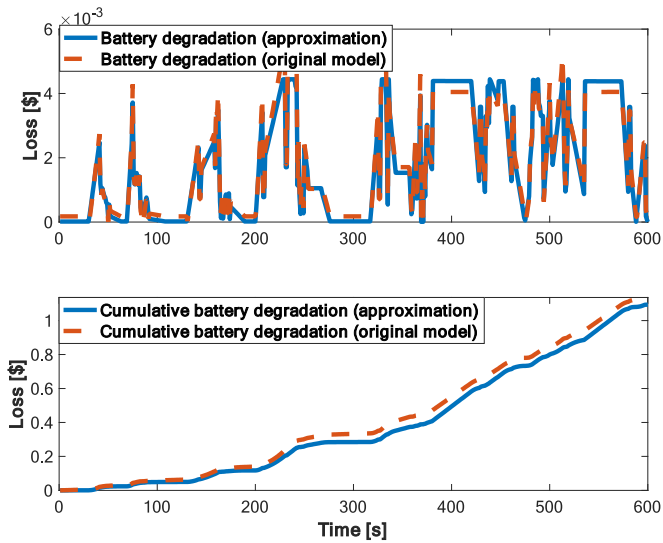

**Fig. 11.** Comparison of battery degradation: approximation vs original Model.

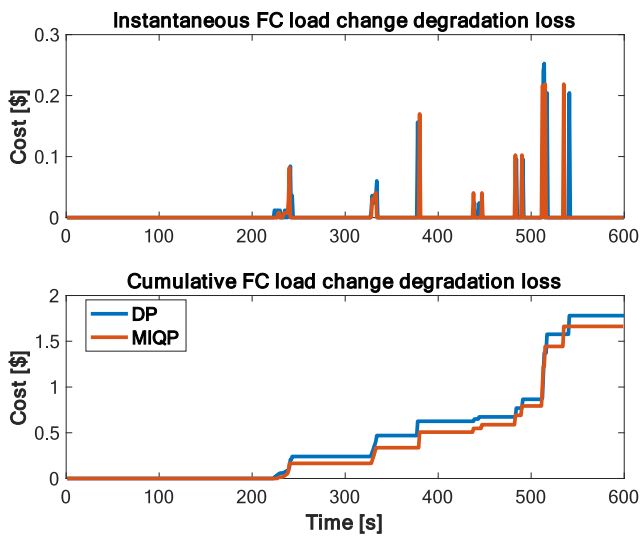

**Fig. 12.** The FC load change degradation losses resulting from the DP and proposed methods with Collective Stack Control.

practical operating range, thereby maintaining the accuracy of battery SOC updates.

A fine grid size can yield near-optimal results, albeit at the cost of slower computational speed. On the other hand, a coarse grid size may lead to suboptimal outcomes. The relatively coarse grid size of the FC supply power constrains the minimal change in FC supply power, resulting in elevated FC load change loss, as shown in Fig. 12. The coarser grid size also prevents the DP from consistently operating within the optimal range of the FC efficiency curve, leading to additional $H_2$ consumption cost, as illustrated in Fig. 13.

The computational times for both methods, under the Collective Stack Control, are shown in Table 5. The results demonstrate that the proposed MIQP method is 3 orders of magnitude faster than the DP, making it more feasible for real-time applications. With a 600-s optimization horizon and a 60-s block duration, which is larger than the MIQP computational time (14.38 s), the proposed model ensures real-time control.

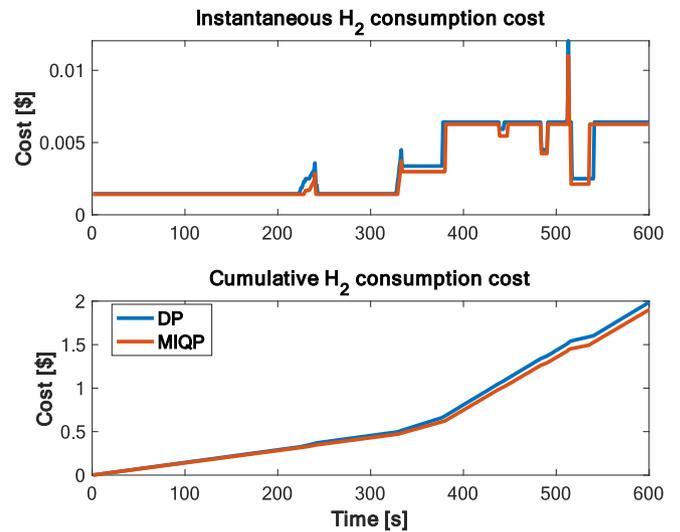

**Fig. 13.** The $H_2$ consumption cost resulting from the DP and proposed methods with Collective Stack Control.

**Table 5**
Comparison of computational times.

| Method | Computational Time |
|---|---|
| MIQP | 14.38 s |
| DP | 32,258.68 s |

### 4.4. Comparative analysis of MIQP method with individual

The power split profiles between FC stacks and the battery pack are depicted in Fig. 14. The FC power is almost always constant relative to the power split profiles in Fig. 9, illustrating the distinctions between ISC and CSC methods. As shown in Fig. 15, the battery SOC behaves differently between the two methods due to the distinct power distribution strategies.

In Fig. 16, the battery degradation cost exhibits no significant differences between the two control methods. In contrast, the total FC power, for the most part, remains steady or changes by only a small amount when using ISC, as demonstrated in Fig. 14. As shown in Fig. 17, compared to the CSC method, the ISC significantly reduces the FC load change degradation loss.

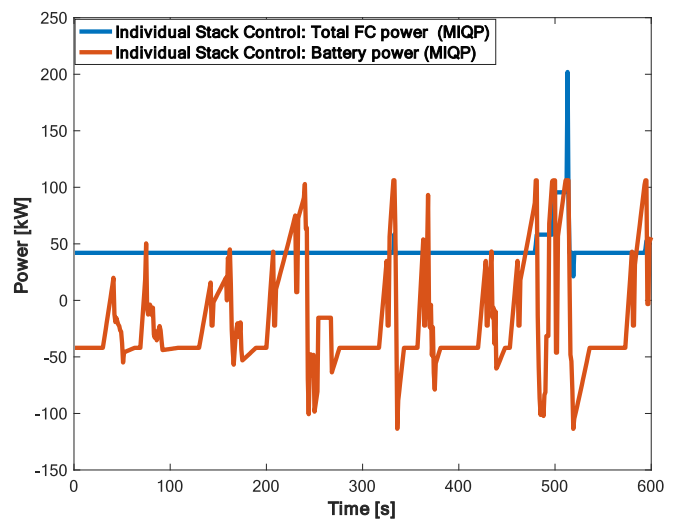

**Fig. 14.** Power split profiles using the proposed method and ISC.

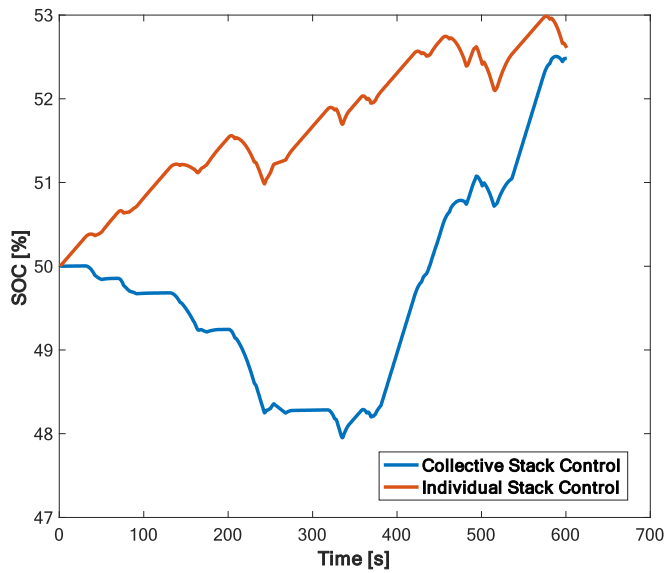

**Fig. 15.** Comparison of battery SOC between CSC and ISC methods.

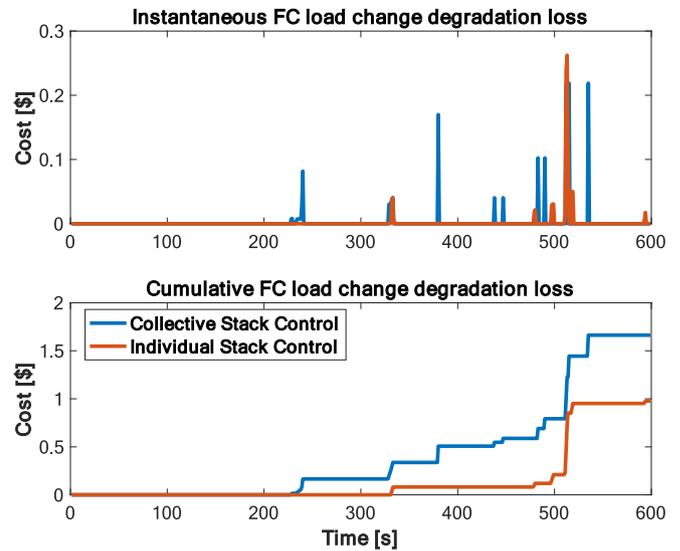

**Fig. 17.** Comparison of FC load change degradation costs between CSC and ISC methods.

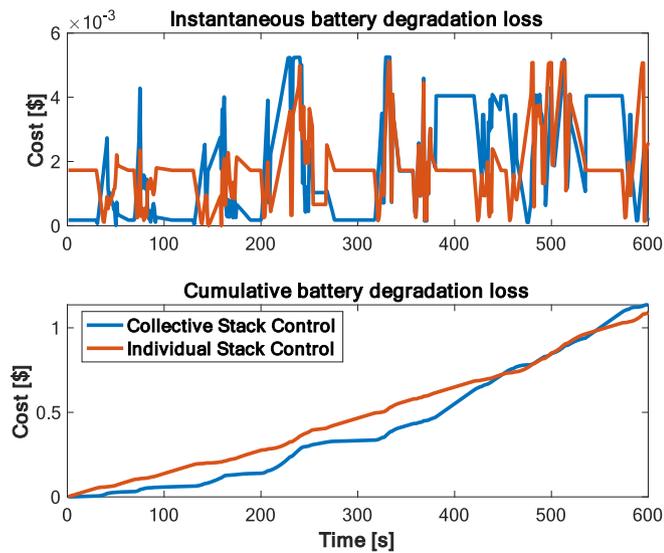

**Fig. 16.** Comparison of battery degradation cost between CSC and ISC methods.

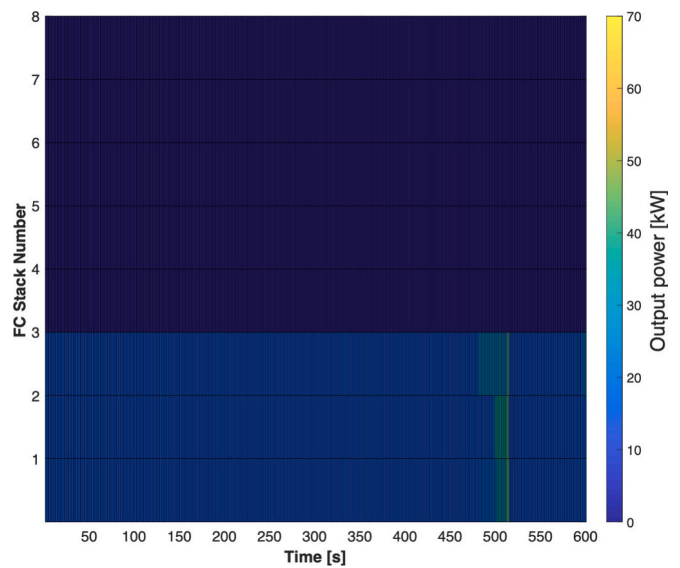

**Fig. 18.** Power split (heat map) among different FC stacks resulting from the ISC method.

Given the involvement of multiple FC stacks, a heatmap of output powers for each FC stack is introduced in Fig. 18, to illustrate the differences in supply power among the stacks. It clearly shows that only stacks 1, 2, and 3 provide power, while stacks 4 to 8 remain inactive, contributing zero power. This depiction clearly shows the variability in the output power of each FC stack at the same point in time. With the Individual Stack Control method, the system shuts down the additional FC stack to ensure that the remaining active FC stacks operate within a high-efficiency power range by assigning different supply power to each stack. Notably, stacks 1, 2, and 3 exhibit similar power profiles, typically ranging from 14 to 32 kW, which aligns with the high-efficiency range illustrated in the upper subplot of Fig. 2.

A comparison of fuel consumption costs between Individual Stack Control and Collective Stack Control using the proposed MIQP method is shown in Fig. 19. As a result, controlling each FC stack individually with the proposed EMS leads to improved operational efficiency and lower fuel consumption costs.

In addition to maintaining each FC stack within the high-efficiency range, the ISC also prevents the FC stacks from remaining idle, corresponding with the low-efficiency range. Fig. 20 presents a comparison of FC idling degradation loss between Individual Stack Control and Collective Stack Control using the proposed MIQP method. Whereas the idling degradation loss constitutes the majority of the cost under the Collective Stack Control method, the Individual Stack Control method avoids idle conditions by selectively activating the necessary number of FC stacks.

Out of eight, only three FC stacks (Stack 1, 2 and 3) are on, with the remaining five stacks always off. When a FC stack is off, it experiences no degradation, including idling degradation. The inactive stacks neither consume fuel nor undergo degradation. The active stacks operate within their optimal efficiency range to minimize system-wide degradation.

Additionally, reducing frequent start-stop cycles for inactive stacks helps mitigate start/stop degradation. As a result, stacks that are initially activated remain 'on' throughout the operation, while those that are initially 'off' stay inactive. This is why the on/off cycle is recorded as zero in Table 5. However, since only three FC stacks are actively used, the EMS cannot supply the peak power demand at time 513 s, resulting in additional FC high load degradation loss, as shown in Fig. 21.

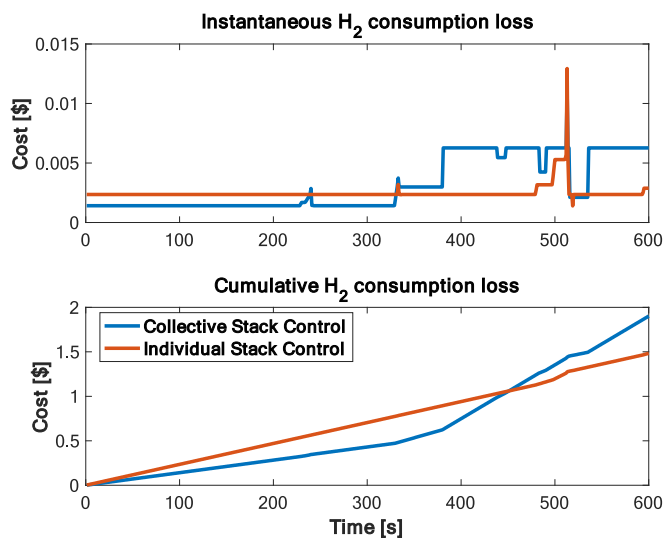

**Fig. 19.** Comparison of fuel consumption costs between CSC and ISC methods.

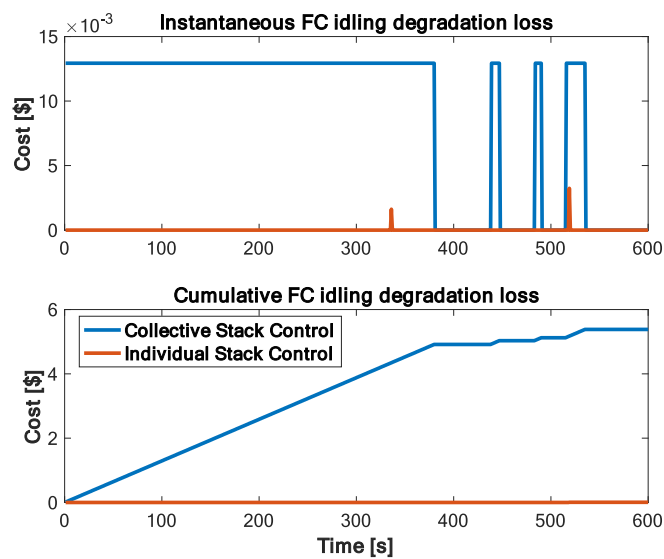

**Fig. 20.** Comparison of FC idling degradation costs between CSC and ISC methods.

As indicated in Table 6, the cumulative total cost of the proposed method, which controls each FC stack individually, is $3.56. This cost is 64.68 % lower than the cost incurred when using the CSC method. These results underscore the benefits of improved system efficiency, enhanced load distribution, and optimal utilization of multiple FC stacks.

The proposed ISC enables tailored configuration of different FC stacks with their specific parameters. This capability could significantly enhance the performance of EMS when different types of FC stacks are deployed in the system. Furthermore, the ISC offers flexibility in system operation and maintenance. It supports consistent monitoring and maintenance of specific stacks, simplifying troubleshooting and stack replacement tasks. This individualized control of FC stacks also strengthens fault tolerance and system resilience [52].

## 5. Conclusion

This study proposed an MIQP-based and health-aware EMS for multiple stack hydrogen FC and battery hybrid systems using a novel ISC

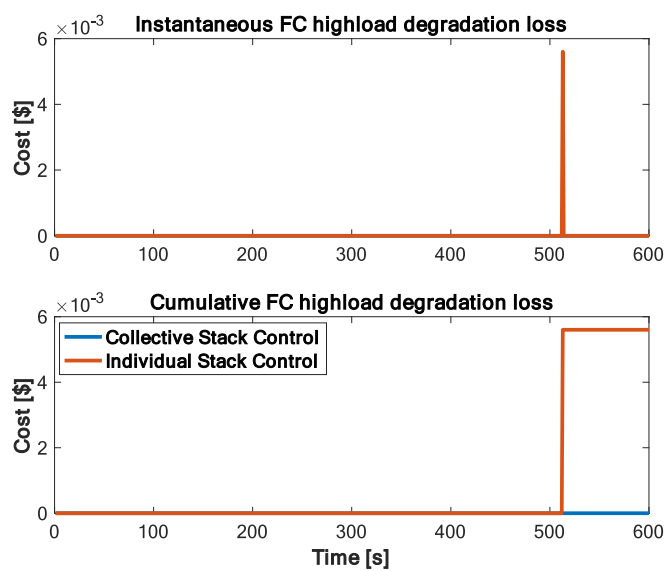

**Fig. 21.** Comparison of high-power load FC degradation costs between CSC and ISC methods.

**Table 6**
Summary of the costs crossing different control methods.

| Cost ($) | CSC | | ISC |
|---|---|---|---|
| | DP | MIQP | MIQP |
| Battery degradation loss | 1.1398 | 1.1362 | 1.0898 |
| H2 consumption cost | 1.9870 | 1.9014 | 1.4829 |
| FC idling loss | 5.3410 | 5.3798 | 0.0048 |
| FC high load loss | 0 | 0 | 0.0056 |
| FC load change loss | 1.7802 | 1.6626 | 0.9768 |
| FC on/off switch loss | 0 | 0 | 0 |
| Total cost (USD) | 10.24 | 10.08 | 3.56 |

strategy that minimize the FC idling time and idling cost. The ISC control strategy offers valuable insights into the advantages of managing each FC stack independently.

The MIQP method is first compared to DP method when using the CSC. The CSC treats all FC stacks as one entity with the same control actions. The comparison shows that the MIQP is comparable with DP's performance, while with a 2243 times faster computation. This efficiency suggests MIQP is effective and beneficial for real-time control of online and complex systems. Based on this, the study reveals that adopting the ISC, which manages each FC stack independently, can substantially enhance the economic performance of the EMS for the hybrid system by reducing the FC idling time and idling cost. This method leads to significant reductions in fuel consumption and costs related to FC and battery degradation, resulting in a total cost reduction of 64.68 % in the examined scenario. It is noted that when power demand is low, the ISC method is even more critical, as the CSC method performs particularly poorly due to FC idling degradation losses.

The proposed MIQP-based EMS, together with the ISC strategy, lays a robust foundation for future exploration and development in the optimization and control of multi-stack FC and their hybrid systems. While this study focuses on a city bus application, the technique could be readily adapted to similar systems, such as trucks and ocean-going vessels [53]. In future studies, we plan to develop more advanced, physics-informed degradation models for fuel cells and batteries, and integrate them with real-world systems for improved fidelity and validation. Furthermore, the optimization results presented in this study are sensitive to specific economic parameters such as fuel price, battery cost, and fuel cell stack cost. Variations in these costs, influenced by economies of scale, technological maturity, market conditions, or differing lifetime expectations, could impact the optimal system configuration. An interesting direction for future research is the development of a more robust optimization framework that explicitly accounts for long-term economic parameter variations and aims to minimize overall lifecycle operation costs.

### CRediT authorship contribution statement

**Junzhe Shi:** Writing – review & editing, Writing – original draft, Visualization, Validation, Software, Methodology, Investigation, Formal analysis, Data curation, Conceptualization. **Ulf Jakob Flø Aarsnes:** Writing – review & editing, Writing – original draft, Supervision, Methodology, Funding acquisition, Conceptualization. **Shengyu Tao:** Conceptualization, Validation, Visualization, Writing – original draft, Writing – review & editing. **Ruiting Wang:** Writing – review & editing, Writing – original draft. **Dagfinn Nærheim:** Writing – review & editing, Investigation, Funding acquisition, Data curation. **Scott Moura:** Writing – review & editing, Writing – original draft, Supervision, Investigation, Funding acquisition, Conceptualization.

### Declaration of competing interest

The authors declare that they have no known competing financial interests or personal relationships that could have appeared to influence the work reported in this paper.

### Acknowledgement

This work was supported in part by NORCE under grant number 336527.

### Data availability

Data will be made available on request.